\documentclass[a4paper,fleqn,usenatbib]{mnras}

\usepackage{txfonts}

\usepackage[T1]{fontenc}
\usepackage{ae,aecompl}

\usepackage{graphicx}	 		
\usepackage{amssymb}   		 
\usepackage{nicefrac}	  		
\usepackage[table]{xcolor}    

\title[Solar System Planet Transit Zones]{Transit Visibility Zones of the Solar System Planets}

\author[R. Wells et al.]{
R. Wells$^{1}$\thanks{E-mail: \href{mailto:rwells02@qub.ac.uk}{rwells02@qub.ac.uk}},
K. Poppenhaeger$^{1,2}$,
C. A. Watson$^{1}$ and
R. Heller$^{3}$
\\
$^{1}$Astrophysics Research Centre, Queen's University Belfast, Belfast BT7 1NN, UK
\\
$^{2}$Harvard-Smithsonian Center for Astrophysics, 60 Garden Street, Cambridge, MA 02138, USA
\\
$^{3}$Max Planck Institute for Solar System Research, Justus-von-Liebig-Weg 3, 37077 G\"ottingen, Germany
}

\date{Accepted XXX. Received YYY; in original form ZZZ}

\pubyear{2017}

\begin{document}
\label{firstpage}
\pagerange{\pageref{firstpage}--\pageref{lastpage}}
\maketitle

\begin{abstract}
The detection of thousands of extrasolar planets by the transit method naturally raises the question of whether potential extrasolar observers could detect the transits of the Solar System planets. We present a comprehensive analysis of the regions in the sky from where transit events of the Solar System planets can be detected. We specify how many different Solar System planets can be observed from any given point in the sky, and find the maximum number to be three. We report the probabilities of a randomly positioned external observer to be able to observe single and multiple Solar System planet transits; specifically, we find a probability of 2.518\% to be able to observe at least one transiting planet, 0.229\% for at least two transiting planets, and 0.027\% for three transiting planets. We identify 68 known exoplanets that have a favourable geometric perspective to allow transit detections in the Solar System and we show how the ongoing K2 mission will extend this list. We use occurrence rates of exoplanets to estimate that there are $3.2\pm1.2$ and $6.6^{+1.3}_{-0.8}$ temperate Earth-sized planets orbiting GK and M dwarf stars brighter than $V=13$ and $V=16$ respectively, that are located in the Earth's transit zone.

\end{abstract}


\section{Introduction}{\label{sec:intro}}

Over the past decade, the number of known of exoplanets has grown immensely. Over 3500 exoplanets have been discovered to date\footnote{\url{http://exoplanet.eu} \citep{schneider2011}}, going from close-orbiting, Jupiter-sized planets (hot Jupiters) to those similar to Earth in mass and size, e.g. Kepler-78b \citep{pepe2013}. A number of temperate extrasolar planets have now been discovered \citep{vogt2010lick,anglada2013dynamically,jenkins2015discovery}, including one around our nearest neighbour, Proxima Centauri \citep{anglada2016terrestrial}.

The habitable zone (HZ) is defined as the region around a star where a terrestrial planet could sustain liquid water. It largely depends on the stellar flux received and surface atmospheric pressure, along with other planetary and stellar parameters \citep{Kasting1993, kopparapu2013habitable}. That said, planets and even moons beyond the HZ could still have temperate regions, e.g. if they are subject to tidal heating \citep{Reynolds1987, BarnesHeller2013}. \citet{2015ApJ...807...45D} found the occurrence rate of Earth-sized ($0.5-1.4\,R_\oplus$) planets within the HZ around M dwarfs to be $0.68^{+0.13}_{-0.08}$ planets per star, while \citet{2013PNAS..11019273P} found that $22\pm8\%$ of GK stars harbour slightly larger planets ($1-2\,R_\oplus$) in their HZ. These values were calculated with the HZ defined as the region around the star which receives between 0.25 and 4 times the incident stellar flux of Earth. The 39\footnote{\url{http://phl.upr.edu/projects/habitable-exoplanets-catalog} (Planetary Habitability Laboratory, University of Puerto Rico at Arecibo)} potentially habitable planets discovered thus far plus the expected yield of the PLATO mission \citep{rauer2014plato} naturally makes us wonder whether in fact some of these worlds could be inhabited, possibly even by intelligent beings.

The vast majority of exoplanets are currently detected through transits. Conversely, one can ask the question where in the Milky Way the transits of the Solar System planets can be observed. These regions are of particular interest to us, as any potential civilisation that would detect transiting planets around the Sun might have a reason to try and send us deliberate messages in trying to establish contact. This argument, in turn, could serve as a guide for our own search for extraterrestrial intelligence (SETI) \citep{1988ATsir1531...31F,2004IAUS..202..445C,2008AAS...212.0406C,nussinov2009some,heller2016}.

\begin{figure*}
	\begin{center} 
		\includegraphics[width=5in]{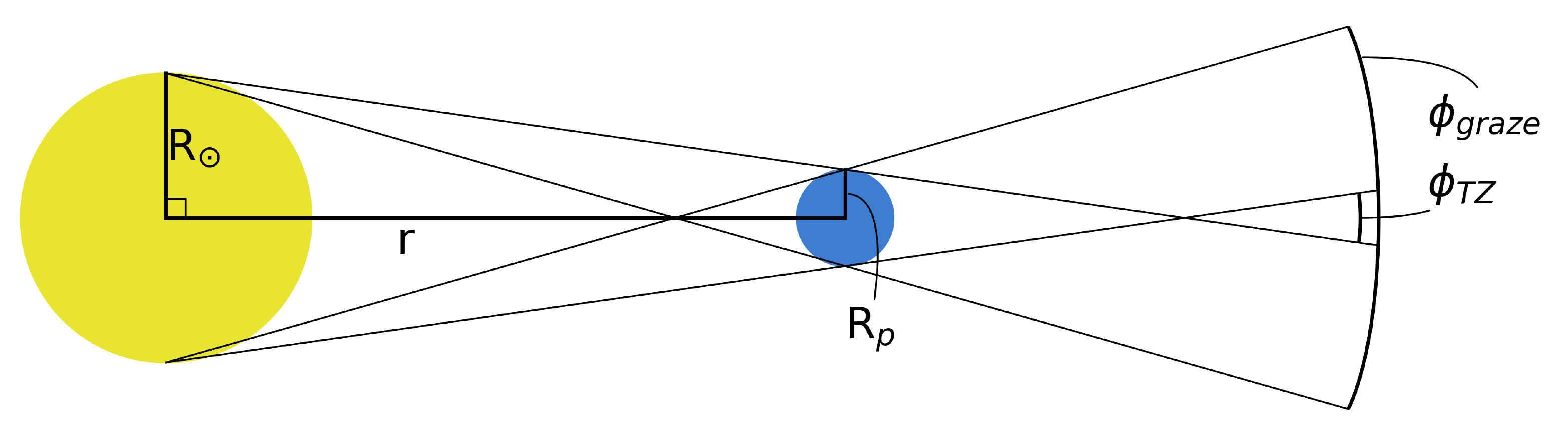} 
		\caption{The transit zone of a planet (centre) projected out from the Sun (left). Sizes and distance not to scale. \label{fig:transitzone}} 
	\end{center} 
\end{figure*}

\section{Methods}{\label{sec:transitzones}}

\subsection{Definition of Transit Zones}{\label{sec:deftransitzones}}

A Transit Visibility Zone -- or simply Transit Zone -- is the projection of a planet onto the celestial plane, from where it is possible to detect transits of the planet in front of the Sun. Fig.~\ref{fig:transitzone} shows the geometry and construction of a Solar System planet's transit zone. Using trigonometry, $\phi_{\rm TZ}$, the angle in which an observer would see a full transit and $\phi_{\rm graze}$, the angle to observe a grazing transit, are given by \citep{heller2016}

\begin{equation} 
\label{eq:phi_etz}
\phi_{\rm TZ} = 2 \left( \arctan \left( \frac{R_{\odot}}{r} \right) - \arcsin \left( \frac{R_{p}}{\sqrt{(r^2 + R_{\odot}^2)}} \right) \right)\,,
\end{equation} 

\begin{equation}
\label{eq:phi_graze}
\phi_{\rm graze} = 2 \arctan \left( \frac{(R_{p} + R_{\odot})}{r} \right)\,,
\end{equation}
where $R_{\odot}$ is the radius of the Sun, $R_{p}$ is the radius of the planet and $r$ is the instantaneous distance between the Sun and the planet, which varies over the course of the planet's orbit if it has non-zero eccentricity. This distance can be written as $r = a(1-e^2)/(1+e \cos(\theta))$ where $a$ is the semi-major axis, $e$ is the eccentricity and $\theta$ is the true anomaly - the angle between the planet's current position and the point in the orbit when it is closest to the Sun.

We note that $\phi_{\rm TZ}$ is commonly approximated by $2R_{\star}/a$ \citep{Borucki1984}. As we show below, this works well for small planets at large separations. However, geometric transit probabilities obtained for larger Solar System planets differ to the results obtained via Equation~(\ref{eq:phi_etz}) by up to 21\% (see Section~\ref{sec:difftoapprox}). We note that this approximation is widely used in the community even when it does not hold, in particular for exoplanets which have a small semi-major axis including those orbiting late-M dwarfs and hot Jupiters. For example, TRAPPIST-1g \citep{gillon2017seven}, a temperate Earth-sized planet orbiting an M8 star gives a 10\% difference between Equation~(\ref{eq:phi_etz}) and the approximation, which is close to the largest disagreement in the Solar System (Jupiter, 11\%). This therefore has an impact on the statistics of exoplanetary systems, such as occurrence rate which is computed with the approximation (e.g.\ \citealt{howard2012planet}, \citealt{2015ApJ...807...45D}).

\subsection{Numerical determination of Transit Zones}{\label{sec:numtransitzones}}

In order to calculate the transit zone angles, we collected values of the planetary radii, semi-major axes and the radius of the Sun from the NASA planetary fact sheets\footnote{\url{http://nssdc.gsfc.nasa.gov/planetary/planetfact.html} (Curator: D.R. Williams, NSSDCA)}, which are compiled from recent literature. For the planetary radii, the volumetric mean value was used; i.e. the radius of a sphere with the same volume as the body. 

\begin{figure*}
	\centering
	\includegraphics[width=.9\textwidth]{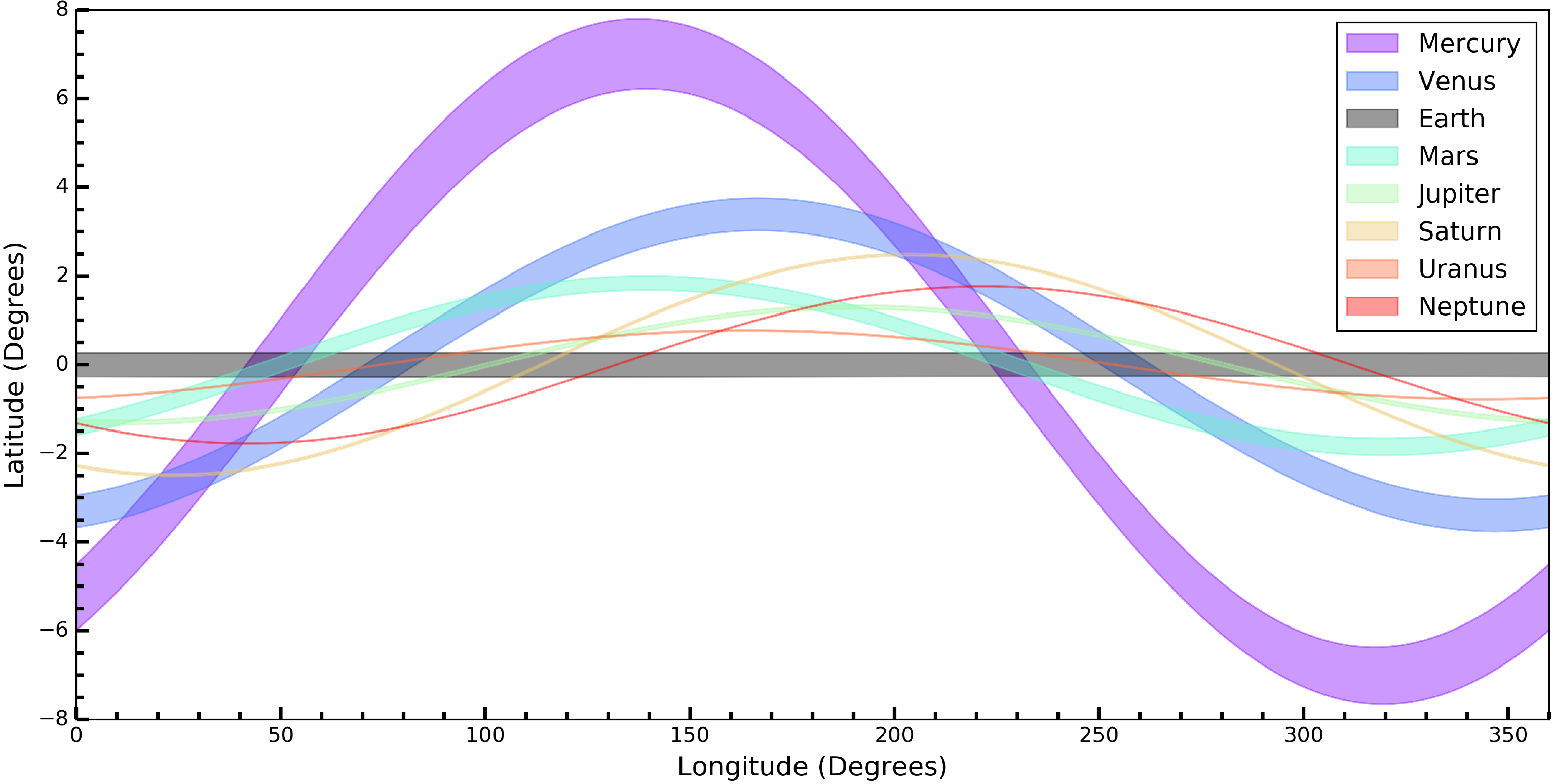}
	\caption{Projections of the transit visibility zones (non-grazing) of all the Solar System planets in heliocentric ecliptic coordinates. Each planet's transit zone is defined as a coloured bounded region.}
	\label{fig:allhelioecliptic}
\end{figure*}

The transit zone and grazing angles ($\phi_{\rm TZ}$ and $\phi_{\rm graze}$ in Fig.~\ref{fig:transitzone}) for the Solar System planets are given in Table~\ref{tab:phi&depth} computed using Equations~(\ref{eq:phi_etz}) and (\ref{eq:phi_graze}). We also list the expected transit depths $\Delta F=(R_{\rm p}/R_\odot)^2$. The relative differences between the angles of the full and grazing transit zones, calculated as per $\Delta\phi=\phi_{\rm graze}/\phi_{\rm TZ}-1$, are also given. Variations are largest for large planets with small transit zones, such as Jupiter and Saturn. These differences are discussed more in detail in Section~\ref{sec:res&disc}. Table~\ref{tab:phi&depth} also serves as a comparison of how detectable each planet is, where a larger $\phi_{\rm TZ}$ suggests more possible extrasolar observers of the respective planetary transits and $\Delta F$ implies easier detection of the transit feature.

\begin{table}
	\centering
	\caption{The transit zone angles and transit depths of the Solar System planets.}
	\begin{tabular}{lcccc}
		\hline\hline
		Planet & $\phi_{\rm TZ}$ (\degr) & $\phi_{\rm graze}$ (\degr) & $\Delta \phi$ (\%) & $\Delta F$ (\%) \\
		\hline
		Mercury & 1.3714 & 1.3810 & 0.7 & 0.0012 \\
		Venus & 0.7301 & 0.7429 & 1.8 & 0.0076 \\
		Earth & 0.5279 & 0.5376 & 1.8 & 0.0084 \\
		Mars & 0.3479 & 0.3513 & 1.0 & 0.0024 \\
		Jupiter & 0.0921 & 0.1127 & 22.4 & 1.0104 \\
		Saturn & 0.0512 & 0.0605 & 18.3 & 0.7010 \\
		Uranus & 0.0268 & 0.0288 & 7.6 & 0.1330 \\
		Neptune & 0.0171 & 0.0183 & 7.3 & 0.1253 \\
		\hline
	\end{tabular}
	\label{tab:phi&depth}
\end{table}

To calculate the positions of the transit zones on the celestial plane, we obtained orbital data from JPL HORIZONS\footnote{\url{http://ssd.jpl.nasa.gov/horizons.cgi} \citep{giorgini1996jpl}}, where a single complete orbit of each planet was taken in the heliocentric ecliptic coordinate system. In heliocentric ecliptic coordinates, Earth's transit zone is a roughly $0.53^\circ$ wide disk of constant latitude at ecliptic latitude $b=0$ (see \citealt{heller2016}). The other Solar System planets describe sinusoidal tracks determined by their Keplerian orbital elements. In order to make identifications of overlapping transit visibility zones more convenient, those discrete numerical values were fit with functions of the form

\begin{equation}
\label{eq:sinfunction}
l = A \sin(b - d)\,,
\end{equation}
where $l$ is the longitude, $b$ is the latitude. $A$ is the planet's orbital inclination with respect to the ecliptic and $d$ is related to the planet's argument of the periastron. We treated both $A$ and $d$ as are free parameters in our fitting procedure. The upper and lower bounds of the transit visibility zones were then created by adding and subtracting the value of $(\phi_{\rm TZ}/2)$ at each point along the orbit, respectively. This is a slight approximation; the exact transit zone borders would be derived by converting to planet-ecliptic coordinates before adding/subtracting the angle. Therefore by neglecting this step, the thickness of the transit zones are marginally thinner towards $b=0$. We computed the minimum (at $b=0$) and maximum (at maximum latitude $l_{\rm peak}$ of a given planet in ecliptic coordinates) distances between the upper and lower boundaries and then by comparing the difference between these distances, we estimated the change in thickness of the zones. Due to the small angles involved, we find this is at most an effect of 0.7\% in thickness for Mercury, less than 0.2\% for the other planets and no effect for Earth as the ecliptic coordinate system is coplanar with Earth's orbit. We therefore decided to work with the approximated transit visibility zones. 

The transit zones were computed using only one orbit of data which we have assumed to be stationary. However, the shapes and positions of the transit zones would change slightly over the course of many orbits due to orbital precession from planet-planet interactions, the motion of the Sun from interacting with Jupiter and any planet-moon interactions \citep{heller2016}. We calculated how long the transit zones would be valid by taking values\footnote{\url{https://ssd.jpl.nasa.gov/txt/p_elem_t1.txt}} for the change in longitude of perihelion from \citet{standish1992orbital} (roughly $0.3^{\circ}$ per century, excluding Venus) which we assume to be constant and comparing them to half the width of the transit zones. From this we found that the transit zones of the terrestrial planets would be valid for thousands of years, while the smaller transit zones of the Jovian planets would be valid for hundreds of years. However, Venus precesses two orders of magnitude slower than the other planets and therefore its transit zone would be valid for a few hundred thousand years. We also note that stars are not fixed in the sky and instead have proper motions of around 1 arcsecond/yr for nearby stars ($<10$ pc) and around 100 mas/year for more distant stars ($>20$ pc). This means that nearby and more distant stars centred in the transit zone of Earth would leave the zone after roughly 1,000 and 10,000 years respectively if they were moving perpendicular to the boundaries. These timescales are 3 times longer for traversing the transit zone of Mercury and 30 times shorter for the case of Neptune, simply by comparing the relative sizes of $\phi_{\rm TZ}$.


Timescales for the overlap regions of transit zones are more complex due to their various sizes and the movement of multiple transit zones. We give one example here, for Mars-Jupiter, which consists of two areas with width of ${\sim} 14^{\circ}$. We took the central point of one of the areas and then precessed the transit zones through time until the central point at $t=0$ was no longer inside the overlap region. We found that the transits of Mars and Jupiter would both be observable for approximately 1700 years. We also find that nearby and more distant stars would cross the nearest overlap region border in 140 and 1,400 years respectively if moving perpendicular to the border. Other overlapping regions will be observable for different amounts of time due to their various sizes and mutual precession speeds. The code to produce the tables and figures in this work is available on GitHub\footnote{\url{https://github.com/ExoRob/Transit-Zones}}.

\section{Results}{\label{sec:res&disc}}

\subsection{Location of Transit Zones in the sky}

We show the locations of the transit zones of the Solar System planets in Fig.~\ref{fig:allhelioecliptic}. All sets of the planets' transit zones were searched for mutual overlaps from where multiple planets could be detected. Observers here would see our Solar System similarly to how we see transiting multi-planet systems. It can be seen from Fig.~\ref{fig:allhelioecliptic} that the maximum number of different transits visible from any one point in the sky is three. This shows that exoplanetary systems with multiple transiting planets may still hide further companions, as highlighted by \citet{2013ApJ...762..129K}. Different sets of three planets can be observed to transit from different points in the sky, which we find in agreement with \citet{brakensiek2016efficient}. 

\begin{figure*}
	\centering
	\includegraphics[width=6in]{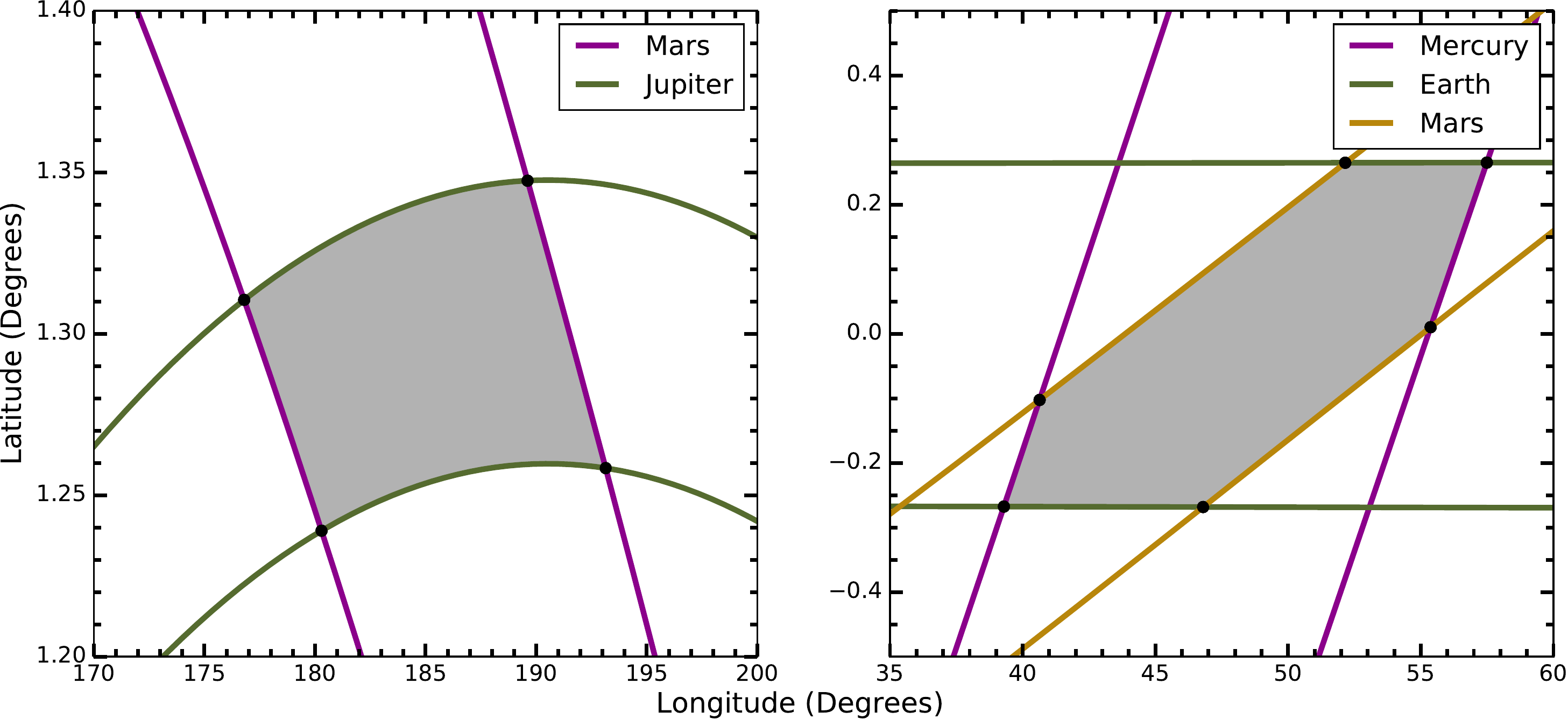}
	\caption{Examples of regions where the non-grazing transit zones intersect, given in ecliptic coordinates. The black circles are the points where the lines intersect -- equatorial coordinates of these points are given in Appendix~\ref{app:regions} for all overlap regions.}
	\label{fig:regionssubplot}
\end{figure*}

Specifically, we find 28 two-planet and 8 three-planet sets, each of which comprises two overlapping regions in opposite directions on the celestial plane that are offset by roughly 180 degrees of longitude, except for the Mercury-Earth-Uranus set. Fig.~\ref{fig:regionssubplot} shows a zoom into one region of Mars-Jupiter and one region of the Mercury-Earth-Mars overlapping regions to give an idea of the shapes and sizes of these areas. All sets of planets where overlap exist are given in Appendix~\ref{app:regions} with equatorial coordinates of the intersection points, converted using the SkyCoord Astropy Python package \citep{robitaille2013astropy} with a distance of 30\,ly. A distance is required due to the origin shift when converting between heliocentric and geocentric coordinate systems. 30\,ly was chosen because stars of interest are typically farther away than ca.\ 10~pc, and assuming greater distances did not change computed coordinates past $10^{-5}$ degrees ($0.04''$). The coordinates are provided to be used as a quick check as to whether any given star falls into a transit zone.

\subsection{Probabilities to observe Solar System planet transits}

By comparing the solid angle covered by each transit zone to the area of the whole sky ($4\pi$), it is possible to obtain the probabilities of each planet to be transiting for a randomly positioned extrasolar observer. The probabilities of each overlap region were also found in the same manner, and are given in Appendix~\ref{app:probabilities}, both as the absolute probability and as a factor of the probability to detect Earth. For individual planets, the area of the transit zones were calculated by using the trapezoidal rule to find the area beneath each of the boundaries. Subtracting the area below the lower boundary from the upper boundary gave the area of the transit zone. For the two- and three-planet overlap region cases, first the curves defining the boundaries of each region were found. Then the areas were found the same way as in the single planet case. Adding these areas together then gave the total area for the set of planets. The areas for all cases were then converted to steradians, and divided by $4\pi$ to return geometric transit probabilities. 

The probabilities of a randomly positioned extrasolar observer being able to observe at least one, two, or three transits were calculated by adding up the individual transit probabilities listed in Table~\ref{app:probabilities}, giving

\begin{itemize}
\item $P_\mathrm{1} = 2.518\%$ to observe at least one transiting planet;

\item $P_\mathrm{2} = 0.229\%$ to observe at least two transiting planets; and

\item $P_\mathrm{3} = 0.027\%$ to observe three transiting planets.
\end{itemize}

These probabilities were calculated simply from Table~\ref{tab:probabilities} and are therefore only applicable to observing the Solar System. Exoplanetary systems will have different coplanarity and multiplicity which will consequently lead to different probabilities. Using the values given in Table~\ref{app:probabilities}, we then calculate the probability of an observer inside the Earth's transit zone to be able to see other Solar System planets in transit as well. We find that an observer viewing transits of Earth would have a 24\% chance of detecting at least one more planet in the Solar System via transits, purely from a geometric point of view.

\begin{figure}
	\centering
	\includegraphics[width=\columnwidth]{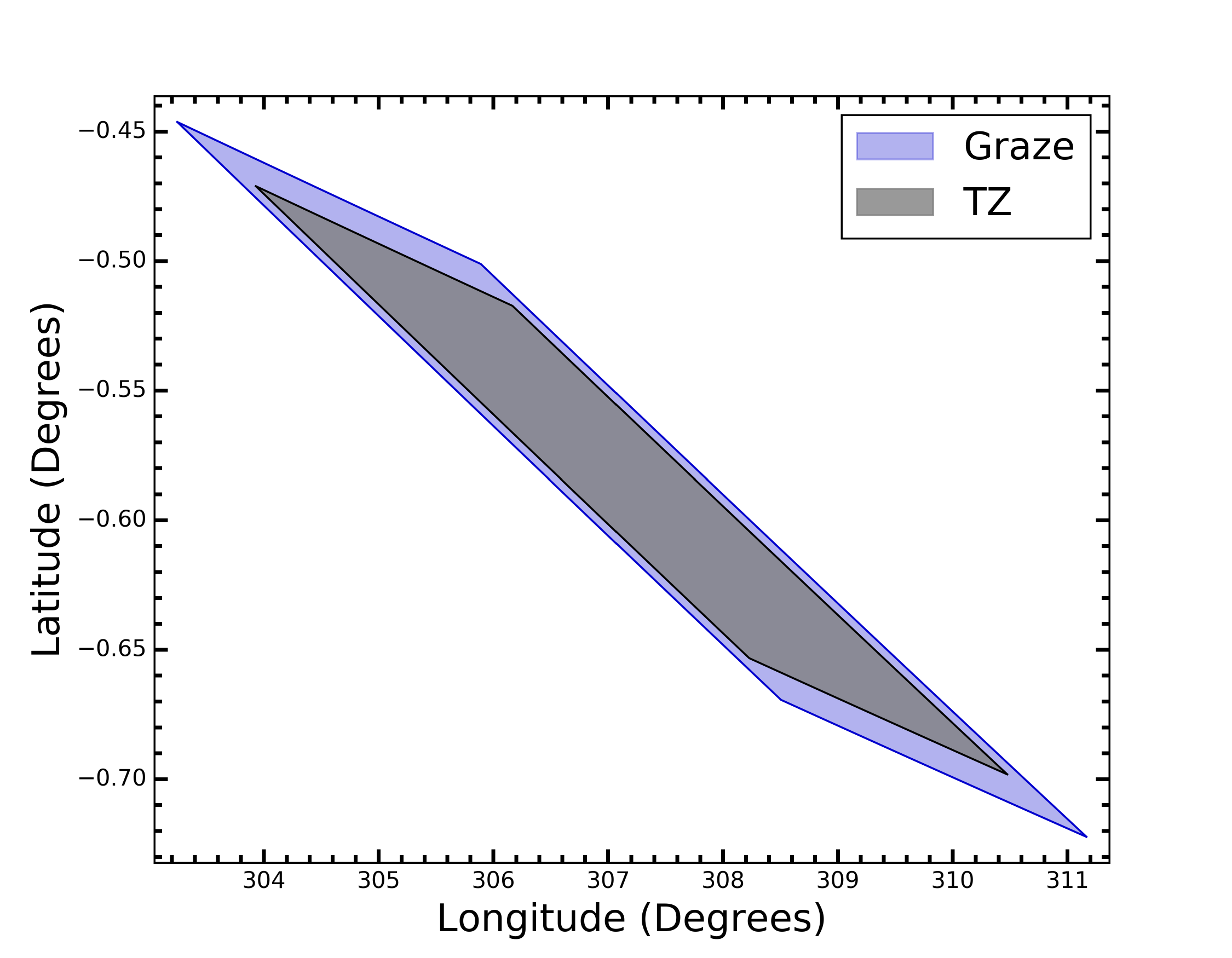}
	\caption{Comparison of the regions where grazing transits (blue) and full transits (grey) of both Jupiter and Saturn could be observed, given in heliocentric ecliptic coordinates. The area of the blue region is approximately 45\% larger than the area of the grey region.}
	\label{fig:graze_tz_comparison}
\end{figure}

The regions where an extrasolar observer could detect grazing transits of multiple planets are significantly larger for some sets. This is simply due to the grazing angle being larger than the transit zone angle (see Table~\ref{tab:phi&depth}), in particular for cases involving the gas giants. This effect is also more apparent for the Jovian planets because their transit zones are smaller due to their semi-major axes being much larger than those of the terrestrial planets. For example, the grazing region between Jupiter and Saturn can be seen in Figure~\ref{fig:graze_tz_comparison}, which is 45\% larger compared to the full transit zone of these 2 planets. Therefore, searches can be extended by the inclusion of regions where grazing transits can be detected and would give significantly more area of the sky for sets that include Jovian planets. 

\begin{figure*}
	\centering
	\includegraphics[width=\textwidth]{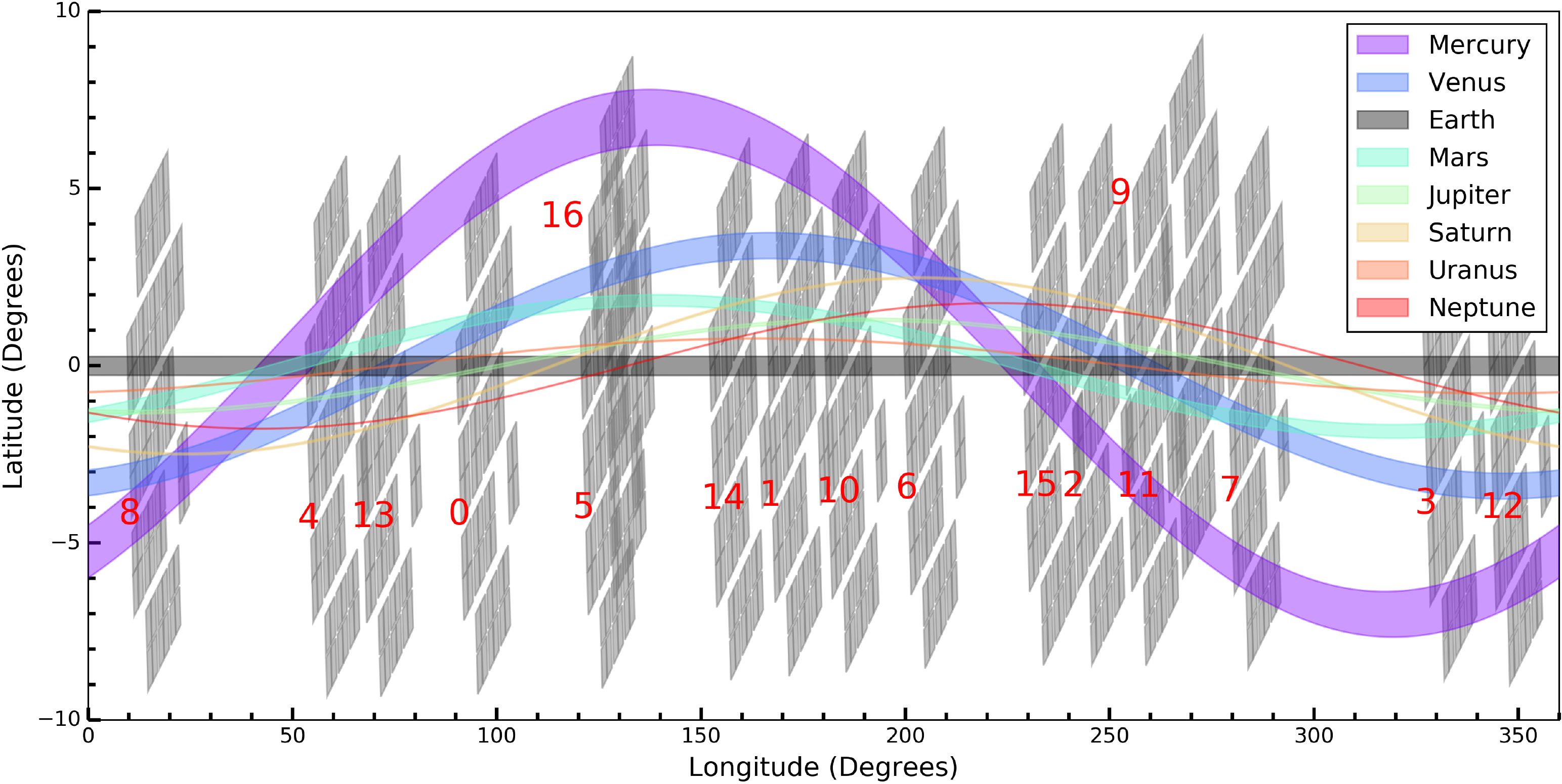}
	\caption{As Fig.~\ref{fig:allhelioecliptic}, with the fields of the K2 mission overlaid. The campaign number is given in red for each field.}
	\label{fig:allecliptic+k2fields}
\end{figure*}

\subsection{Differences to approximate calculations}{\label{sec:difftoapprox}}

Table~\ref{tab:propcomp} shows a comparison of probabilities computed using the transit zone angle to those calculated using an estimate of $2R_{\sun}/a$, an approximation originally proposed by \citet{Borucki1984}. The values we obtain for the terrestrial planet overlap regions match well with the approximation, whereas the Jovian planet overlap regions differ by up to 21\%. The Mercury-Earth-Uranus region shows an extremely large discrepancy between the approximation and the exact calculation; this is due to the region's small size, where small changes in the transit zone thickness change the area by relatively large amounts. The discrepancies for the other sets of planets are due to the approximation treating planets as point objects, i.e. ignoring the planetary radius. Therefore the approximation significantly overestimates transit probabilities for overlap regions that include the giant planets, as seen in Table~\ref{tab:propcomp}. 

\begin{table}
	\centering
	\caption{Comparison of probabilities computed using $\phi_{\rm TZ}$ and the approximation, \nicefrac{2R}{a}. A full table of results using $\phi_{\rm TZ}$ is given in Appendix~\ref{app:probabilities}.}
	\begin{tabular}{lccc}
		\hline\hline
		Set of planets & P$(\phi_{\rm TZ})$ & P$(\nicefrac{2R}{a})$ & $\Delta$P (\%) \\
		\hline
		Mercury, Mars & $2.73 {\times} 10^{-4}$ & $2.75 {\times} 10^{-4}$ & -0.9 \\
		Venus, Earth & $3.16 {\times} 10^{-4}$ & $3.22 {\times} 10^{-4}$ & -1.8 \\
		Earth, Mars & $2.80 {\times} 10^{-4}$ & $2.84 {\times} 10^{-4}$ & -1.4 \\
		Earth, Jupiter & $1.05 {\times} 10^{-4}$ & $1.17 {\times} 10^{-4}$ & -12.2 \\
		Jupiter, Saturn & $1.05 {\times} 10^{-5}$ & $1.27 {\times} 10^{-5}$ & -21.3 \\
		Uranus, Neptune & $8.44 {\times} 10^{-7}$ & $9.08 {\times} 10^{-7}$ & -7.6 \\
		Mercury, Venus, Saturn & $3.21 {\times} 10^{-5}$ & $3.52 {\times} 10^{-5}$ & -9.5 \\
		Mercury, Venus, Neptune & $1.60 {\times} 10^{-6}$ & $1.74 {\times} 10^{-6}$ & -8.4 \\
		Mercury, Earth, Mars & $2.08 {\times} 10^{-4}$ & $2.10 {\times} 10^{-4}$ & -1.2 \\
		Mercury, Earth, Uranus & $3.89 {\times} 10^{-8}$ & $8.78 {\times} 10^{-8}$ & -125.6 \\
		Mercury, Mars, Uranus & $4.62 {\times} 10^{-6}$ & $4.88 {\times} 10^{-6}$ & -5.7 \\
		Venus, Earth, Uranus & $2.08 {\times} 10^{-5}$ & $2.18 {\times} 10^{-5}$ & -4.7 \\
		Mars, Jupiter, Neptune & $2.19 {\times} 10^{-6}$ & $2.55 {\times} 10^{-6}$ & -16.8 \\
		Jupiter, Saturn, Uranus & $1.96 {\times} 10^{-6}$ & $2.22 {\times} 10^{-6}$ & -13.3 \\
		\hline
	\end{tabular}
	\label{tab:propcomp}
\end{table}

\section{Discussion}

\subsection{Exoplanets within the Solar System Transit Zones}

All known exoplanets situated within a transit zone of a Solar System planet are given in Appendix~\ref{app:planetzones}. This list is up-to-date as of March 2017 with information from exoplanet.eu, and contains 65 confirmed and 3 unconfirmed planets. The list is slightly disproportionate due to the inclusion of 43 exoplanets that only fall into Mercury's transit zone. We note, however, that 9 planets have already been found in the Earth's transit zone, though none are expected to be habitable due to their size and stellar radiation received. Of particular interest is EPIC211913977 b (K2-101b, \citealt{mann2017zodiacal}) -- a planet on the border between super-Earth and mini-Neptune from whose position transits of Jupiter, Saturn, and Uranus can be detected. In addition, a hot-Jupiter falls into the transit zones of both Earth and Jupiter and a further 3 planets fall into the overlap region between Venus and Mars. 

The Kepler space telescope will help improve the detection statistics of exoplanets that fall into transit zones. After the second of four reaction wheels failed on the Kepler spacecraft on May 11 2013, the telescope was re-purposed into the ongoing K2 mission \citep{howell2014}. K2 observations entail a series of observing ``campaigns'' of fields that are limited by Sun angle constraints to a duration of approximately 80 days each. Although K2 therefore focuses on detecting short-period exoplanets, this does not rule out finding targets of interest for SETI. Planets in tight orbits have been found in the HZ around M-dwarfs, such as planets e, f and g in the TRAPPIST-1 system, which have periods of 6.06, 9.1 and 12.35 days, respectively \citep{gillon2017seven}. The spacecraft is now kept steady by the radiation pressure from the Sun, which means that the telescope must be aligned with the ecliptic and therefore gives a large amount of targets residing in the transit visibility zones of the Solar System planets. Approximately 4000 targets are observed per campaign in these regions, which corresponds to around 12 transiting exoplanet candidates per campaign assuming the value of $3 {\times} 10^{-3}$ candidates per target from \citet{pope2016transiting}. Fig.~\ref{fig:allecliptic+k2fields} shows all past and finalised future K2 fields at the date of publishing, over-plotted with the transit zones of the Solar System planets from Fig.~\ref{fig:allhelioecliptic}. 

\subsection{Earth analogues in Transit Zones}

Currently no terrestrial exoplanets in HZs around their host stars are known to be located in any of the Solar System planet transit zones. Nevertheless, we can estimate how likely such planets are to occur within a certain volume. 

There are 1022 known G and K dwarf stars listed in the SIMBAD database \citep{wenger2000simbad} that reside in Earth's transit zone with a $V$ magnitude of less than 13, which is the limiting brightness for terrestrial planets in the PLATO mission \citep{rauer2014plato}. It is possible to estimate the number of temperate rocky planets around these stars by approximating the frequency of Earth-size planets within the HZ of Sun-like stars ($\eta_{\earth}$). \cite{batalha2014exploring} evaluated $\eta_{\earth}$ to be $22\pm8\%$ for a radius range of $0.5-1.4 R_{\earth}$ around G and K stars using the results of \citet{2013PNAS..11019273P}, which corresponds to roughly $225\pm82$ Earth analogues in this sample of GK stars in Earth's transit zone.

By combining this with the probability of Earth transiting for a randomly position observer (0.46\% from Table~\ref{tab:probabilities}), this gives a statistical estimate of $1.0\pm0.4$ transiting planets around the known Sun-like stars that resides in Earth's transit visibility zone. The estimate is small due to the low number of known stars out to 13 mag; i.e. the incompleteness of the catalogue. By comparing the number of known GK stars at set magnitudes to the volume increase, we estimated the true number of GK stars with $V<13$ in the Earth's transit zone to be roughly 3150. Using this value instead gives $3.2\pm1.2$ transiting temperate Earth-sized planets, where the uncertainty comes from the uncertainty in the value of $\eta_{\earth}$. We note that this is a rather optimistic estimate, due to using boundaries of the ``simple'' habitable zone (incident stellar energy $0.25-4 F_{\earth}$). To date, Kepler-452b \citep{jenkins2015discovery} remains the only exoplanet discovered which comes close to meeting these temperate and Earth-sized requirements; it is a 1.6 R$_{\earth}$ planet which orbits a 13.7 V magnitude G2 star of mass of 1.0 M$_{\sun}$ at a distance of 1.05 AU. However, it does not fall into any Solar System planet transit zone and is unlikely to be rocky \citep{2015ApJ...801...41R,2017ApJ...834...17C}.

We also estimated the expected number of temperate Earth-sized planets orbiting M dwarfs in the Earth's transit zone using the same method used for GK stars. To do this we took a system of an M3 star with an Earth-sized planet orbiting in the centre of the habitable zone. From this we find the probability that a planet will transit to be 1.44\%, using values for an M3 star from \citet{2009ApJ...698..519K}. We take the limiting $V$ magnitude for a detection to be 16 mag, which gives approximately 673 M dwarfs in the Earth's transit zone after accounting for completeness as before. We use the occurrence rate of $0.68^{+0.13}_{-0.08}$ Earth-sized ($0.5-1.4\,R_\oplus$) planets per M dwarf from \citet{2015ApJ...807...45D}, which again are in the ``simple'' habitable zone for comparison. This gives an expected $6.6^{+1.3}_{-0.8}$ Earth-sized planets in the habitable zone around M dwarf hosts brighter than 16 mag which reside in the Earth's transit zone.

Transit visibility zones of the Solar System planets are particular areas of interest for programs that search for extraterrestrial intelligent life, such as the SETI program (see \citet{tarter2001search} and references therein) that has recently gained momentum by the Breakthrough Listen Initiative \citep{2017PASP..129e4501I}. Intelligent observers in these transit visibility zones  could identify Earth as a habitable world and attempt communication, possibly via beamed radio transmissions. Defining regions of the sky where artificial signals are potentially more likely can cut down the area of the sky required for SETI-type programs to only 0.46\% of the complete sky. Assuming the same sensitivity as the Kepler space telescope, transits of Earth could be detected out to roughly 275\,pc, assuming a signal-to-noise ratio >8 and more than 3 transits -- i.e.\ 3 years or more of continuous data. However, with better technology or data spanning a longer time frame, transits of Earth could be observed out to the edges of the HZ of the galaxy ($\sim$\,1\,kpc from Earth, \citealt{lineweaver2004galactic}). In addition to the references in the introduction, transits of Earth and exoplanets have been discussed in the context of extraterrestrial intelligence \citep{2016MNRAS.459.1233K,2017arXiv170703730F}.

\section{Conclusions}{\label{sec:conclusion}}

We identified the regions of the sky where an observer could detect transits of one or more Solar System planets. The coordinates of these regions have been supplied in Appendix~\ref{app:regions}. The probabilities of detecting transits of a single Solar System planet, as well as set of planets with overlapping transit visibility zones, are given in Appendix~\ref{app:probabilities}. These probabilities have been calculated using the exact equation for the transit angles, which is more precise than those derived using the commonly known approximation of $2R_{\sun}/a$. It is not possible to observe transits of more than three Solar System planets from any extrasolar perspective. Detecting two or three Solar System planets via transits is possible, but rather unlikely (see Appendix~\ref{app:probabilities}). We find a 24\,\% chance for an extrasolar observer with the favourable geometry to see the Earth's solar transit to also observe at least one more Solar System planet in transit.

Sixty-eight of the currently known exoplanets or candidates fall into the transit zones of the Solar System planets Appendix~\ref{app:planetzones}. The list includes nine planets in Earth's transit zone, one planet in a triple-overlap region, and four planets in a double-overlap region. None of these planets, however, are expected to harbour life due to their various physical parameters such as mass and radiation received.

The ongoing K2 mission will detect many exoplanets within the transit zones of the Solar System planets -- we expect K2 to discover approximately 12 exoplanet candidates per campaign in these zones. Although all planets found will be relatively fast-orbiting, this does not rule out habitable planets that orbit around M dwarf stars.

We derived a statistical estimate of three temperate Earth-sized planets in orbits around Sun-like stars that fall in Earth's transit zone, one of which may be detected by the upcoming PLATO mission. Looking at these systems would be similar to seeing the Earth from outside the Solar System, and would be high priority targets for searches for biomarkers and even extraterrestrial intelligence due to the possibility of life and the mutual detection of each others' transits.

\section*{Acknowledgements}

RW thanks the Northern Ireland Department for Education for the award of a PhD studentship. CAW acknowledges support from STFC grant ST/P000312/1. This work was supported in part by the German space agency (Deutsches Zentrum f\"ur Luft- und Raumfahrt) under PLATO grant 50OO1501.

\bibliographystyle{mnras}
\bibliography{paper}

\appendix

\section{Overlap regions of transit visibility zones of Solar System planets}{\label{app:regions}}

\begin{table*}
  \centering
  \caption{Overlap regions of all the Solar System planets, given in equatorial coordinates. Each set has two overlap regions, separated by roughly 180 degrees, with the exception of the Mercury, Earth, and Uranus overlap region. Coordinates are given for both individual overlap regions, with each region's coordinates on a separate row.}
  \fontsize{7.2}{9.2}\selectfont
  \begin{tabular}{lll}
    \hline\hline
 	Planets & RA (Degrees) & Dec (Degrees) \\
   	\hline
    \multicolumn{1}{l}{Mercury, Venus} & 11.7015, 20.1091, 31.2765, 40.6216 & 1.2788, 5.8413, 9.7849, 14.1528 \\
          & 193.1873, 202.1523, 208.7497, 217.5191 & -1.9486, -6.7372, -8.7483, -13.0101 \\ \rowcolor{gray!25}
    \multicolumn{1}{l}{Mercury, Earth} & 36.9870, 41.0814, 50.7379, 55.1542 & 14.3382, 16.1807, 18.2795, 19.8598 \\ \rowcolor{gray!25}
          & 218.9127, 223.0575, 228.5762, 232.7704 & -14.9577, -16.7615, -17.7374, -19.3130 \\
    \multicolumn{1}{l}{Mercury, Mars} & 34.4691, 38.2540, 53.0315, 57.1011 & 13.1559, 14.9191, 19.1176, 20.5116 \\
          & 217.0902, 220.7140, 230.0600, 233.8370 & -14.1317, -15.7547, -18.3091, -19.6934 \\ \rowcolor{gray!25}
    \multicolumn{1}{l}{Mercury, Jupiter} & 29.3177, 30.1135, 43.9977, 44.8553 & 10.6326, 11.0310, 15.6051, 15.9623 \\ \rowcolor{gray!25}
          & 211.1509, 211.9035, 221.7918, 222.5512 & -11.3130, -11.6802, -14.9305, -15.2600 \\
    \multicolumn{1}{l}{Mercury, Saturn} & 19.6025, 20.0142, 33.3310, 33.7779 & 5.5699, 5.7906, 10.7851, 10.9999 \\
          & 200.9720, 201.3855, 211.5137, 211.9181 & -6.1163, -6.3342, -10.1334, -10.3331 \\ \rowcolor{gray!25}
    \multicolumn{1}{l}{Mercury, Uranus} & 35.4333, 35.6512, 50.5541, 50.7928 & 13.6128, 13.7154, 18.2107, 18.3000 \\ \rowcolor{gray!25}
          & 217.4555, 217.6955, 228.1101, 228.3516 & -14.2989, -14.4083, -17.5547, -17.6495 \\
    \multicolumn{1}{l}{Mercury, Neptune} & 25.8266, 25.9521, 38.7173, 38.8517 & 8.8527, 8.9175, 13.3024, 13.3632 \\
          & 207.3073, 207.4350, 216.8692, 216.9966 & -9.3973, -9.4620, -12.7084, -12.7677 \\ \rowcolor{gray!25}
    \multicolumn{1}{l}{Venus, Earth} & 63.9972, 73.6155, 77.2616, 87.1343 & 21.0167, 22.8535, 22.6528, 23.6808 \\ \rowcolor{gray!25}
          & 244.1603, 253.5242, 257.3530, 266.9091 & -21.0507, -22.8362, -22.6683, -23.6680 \\
    \multicolumn{1}{l}{Venus, Mars} & 110.0730, 120.6906, 85.9473, 97.0947 & 23.5889, 22.3728, 24.3262, 24.7999 \\
          & 265.6867, 277.7724, 289.2227, 301.3848 & -24.3019, -24.7984, -23.6588, -22.2724 \\ \rowcolor{gray!25}
    \multicolumn{1}{l}{Venus, Jupiter} & 50.5135, 52.9740, 69.8012, 72.3726 & 17.4529, 18.1865, 21.4512, 21.9065 \\ \rowcolor{gray!25}
          & 230.6284, 232.9169, 249.8003, 252.2384 & -17.4869, -18.1687, -21.4535, -21.8865 \\
    \multicolumn{1}{l}{Venus, Saturn} & 18.4418, 19.7970, 37.7801, 39.2221 & 5.1046, 5.7040, 12.3267, 12.8635 \\
          & 198.4127, 199.7273, 217.8311, 219.1757 & -5.0937, -5.6752, -12.3456, -12.8464 \\ \rowcolor{gray!25}
    \multicolumn{1}{l}{Venus, Uranus} & 67.4052, 68.0222, 84.6971, 85.3428 & 21.7360, 21.8582, 23.4867, 23.5419 \\ \rowcolor{gray!25}
          & 247.4526, 248.0827, 264.6176, 265.2576 & -21.7431, -21.8677, -23.4831, -23.5382 \\
    \multicolumn{1}{l}{Venus, Neptune} & 35.8160, 36.1561, 50.6768, 51.0326 & 12.3626, 12.4928, 16.7364, 16.8447 \\
          & 215.8121, 216.1470, 230.6600, 231.0091 & -12.3614, -12.4897, -16.7322, -16.8386 \\ \rowcolor{gray!25}
    \multicolumn{1}{l}{Earth, Mars} & 33.1817, 44.4152, 49.6725, 61.2482 & 13.0659, 16.6004, 18.5652, 21.0835 \\ \rowcolor{gray!25}
          & 213.7997, 224.2269, 229.7480, 240.9652 & -13.2811, -16.5519, -18.5785, -21.0248 \\
    \multicolumn{1}{l}{Earth, Jupiter} & 111.9535, 116.3318, 86.1188, 90.6865 & 22.1766, 21.5066, 23.1223, 23.1688 \\
          & 266.6015, 271.0545, 291.5136, 295.9531 & -23.1419, -23.1753, -22.2285, -21.5605 \\ \rowcolor{gray!25}
    \multicolumn{1}{l}{Earth, Saturn} & 108.0923, 109.4330, 121.3168, 122.6180 & 22.1264, 21.9660, 20.5977, 20.3351 \\ \rowcolor{gray!25}
          & 288.3799, 289.5843, 301.1709, 302.3523 & -22.1018, -21.9561, -20.6178, -20.3809 \\
    \multicolumn{1}{l}{Earth, Uranus} & 50.2518, 52.3607, 93.4203, 95.7638 & 18.1588, 18.6723, 23.6698, 23.6014 \\
          & 230.6594, 232.8785, 272.6986, 274.9796 & -18.2660, -18.8004, -23.6751, -23.6195 \\ \rowcolor{gray!25}
    \multicolumn{1}{l}{Earth, Neptune} & 124.9518, 125.5307, 142.6482, 143.1965 & 19.2871, 19.1582, 15.0190, 14.8417 \\ \rowcolor{gray!25}
          & 305.2351, 305.8100, 322.4193, 322.9703 & -19.2328, -19.1038, -15.0851, -14.9077 \\
    \multicolumn{1}{l}{Mars, Jupiter} & 10.2013, 13.7149, 356.2880, 359.8240 & 2.9218, 4.5116, -3.0317, -1.4173 \\
          & 177.5839, 180.7676, 189.3645, 192.5865 & 2.4753, 1.0176, -2.5706, -4.0320 \\ \rowcolor{gray!25}
    \multicolumn{1}{l}{Mars, Saturn} & 156.3063, 157.5163, 163.6807, 164.8623 & 11.6364, 11.1583, 8.9726, 8.4712 \\ \rowcolor{gray!25}
          & 335.5695, 336.7211, 344.4525, 345.5888 & -11.8888, -11.4378, -8.6811, -8.1960 \\
    \multicolumn{1}{l}{Mars, Uranus} & 22.7045, 23.8867, 39.7803, 41.0040 & 8.8540, 9.3518, 15.0634, 15.4810 \\
          & 203.6493, 204.9001, 219.0955, 220.4226 & -9.2290, -9.7519, -14.8405, -15.2975 \\ \rowcolor{gray!25}
    \multicolumn{1}{l}{Mars, Neptune} & 357.9897, 358.3683, 6.4707, 6.8470 & -2.2563, -2.0834, 1.2162, 1.3888 \\ \rowcolor{gray!25}
          & 178.6744, 179.0460, 185.8387, 186.2121 & 1.9770, 1.8069, -0.9576, -1.1288 \\
    \multicolumn{1}{l}{Jupiter, Saturn} & 126.3296, 128.8754, 130.6375, 133.1409 & 19.7400, 19.1920, 18.8831, 18.2798 \\
          & 306.3716, 308.6853, 310.8305, 313.1306 & -19.7298, -19.2326, -18.8398, -18.2845 \\ \rowcolor{gray!25}
    \multicolumn{1}{l}{Jupiter, Uranus} & 127.4931, 129.8344, 135.0180, 137.3045 & 19.5875, 19.0687, 17.7098, 17.1015 \\ \rowcolor{gray!25}
          & 307.4337, 309.6116, 315.2746, 317.4065 & -19.6019, -19.1212, -17.6409, -17.0713 \\
    \multicolumn{1}{l}{Jupiter, Neptune} & 175.4018, 176.3556, 180.3649, 181.3143 & 3.3075, 2.9026, 1.2860, 0.8789 \\
          & 0.6033, 1.5658, 355.1585, 356.1218 & -1.1885, -0.7757, -3.4060, -2.9974 \\ \rowcolor{gray!25}
    \multicolumn{1}{l}{Saturn, Uranus} & 129.5076, 130.3481, 131.1154, 131.9524 & 19.1184, 18.9441, 18.7263, 18.5448 \\ \rowcolor{gray!25}
          & 309.6050, 310.3891, 311.0756, 311.8565 & -19.0961, -18.9331, -18.7372, -18.5680 \\
    \multicolumn{1}{l}{Saturn, Neptune} & 76.0659, 77.1420, 79.4563, 80.5380 & 21.3616, 21.4899, 21.6875, 21.7941 \\
          & 256.2442, 257.3057, 259.2907, 260.3563 & -21.3806, -21.5060, -21.6733, -21.7794 \\ \rowcolor{gray!25}
    \multicolumn{1}{l}{Uranus, Neptune} & 158.6035, 159.2136, 159.6183, 160.2270 & 9.7970, 9.5578, 9.4284, 9.1875 \\ \rowcolor{gray!25}
          & 338.6460, 339.2602, 339.5715, 340.1844 & -9.7817, -9.5408, -9.4456, -9.2031 \\
    \multicolumn{1}{l}{Mercury, Venus, Saturn} & 19.6025, 20.0142, 33.3310, 33.7779 & 5.5699, 5.7906, 10.7851, 10.9999 \\
          & 200.9720, 201.3855, 211.5137, 211.9181 & -6.1163, -6.3342, -10.1334, -10.3331 \\ \rowcolor{gray!25}
    \multicolumn{1}{l}{Mercury, Venus, Neptune} & 35.8160, 36.1561, 38.7173, 38.8517 & 12.3626, 12.4928, 13.3024, 13.3632 \\ \rowcolor{gray!25}
          & 215.8121, 216.1470, 216.8692, 216.9966 & -12.3614, -12.4897, -12.7084, -12.7677 \\
    \multicolumn{1}{l}{Mercury, Earth, Mars} & 36.9870, 38.2540, 44.4152, 49.6725, 53.0315, 55.1542 & 14.3382, 14.9191, 16.6004, 18.5652, 19.1176, 19.8598 \\
          & 218.9127, 220.7140, 224.2269, 229.7480, 230.0600, 232.7704 & -14.9577, -15.7547, -16.5519, -18.5785, -18.3091, -19.3130 \\ \rowcolor{gray!25}
    \multicolumn{1}{l}{Mercury, Earth, Uranus} & 50.2518, 50.7379, 50.7928 & 18.1588, 18.2795, 18.3000 \\
    \multicolumn{1}{l}{Mercury, Mars, Uranus} & 35.4333, 35.6512, 39.7803, 41.0040 & 13.6128, 13.7154, 15.0634, 15.4810 \\ 
          & 217.4555, 217.6955, 219.0955, 220.4226 & -14.2989, -14.4083, -14.8405, -15.2975 \\ \rowcolor{gray!25}
    \multicolumn{1}{l}{Venus, Earth, Uranus} & 67.4052, 68.0222, 84.6971, 85.3428 & 21.7360, 21.8582, 23.4867, 23.5419 \\ \rowcolor{gray!25}
          & 247.4526, 248.0827, 264.6176, 265.2576 & -21.7431, -21.8677, -23.4831, -23.5382 \\
    \multicolumn{1}{l}{Mars, Jupiter, Neptune} & 178.6744, 179.0460, 180.3649, 181.3143 & 1.9770, 1.8069, 1.2860, 0.8789 \\
          & 0.6033, 1.5658, 357.9897, 358.3683 & -1.1885, -0.7757, -2.2563, -2.0834 \\ \rowcolor{gray!25}
    \multicolumn{1}{l}{Jupiter, Saturn, Uranus} & 129.5076, 130.3481, 131.1154, 131.9524 & 19.1184, 18.9441, 18.7263, 18.5448 \\ \rowcolor{gray!25}
          & 309.6050, 310.3891, 311.0756, 311.8565 & -19.0961, -18.9331, -18.7372, -18.5680 \\
    \hline
    \end{tabular}
  \label{tab:regions}
\end{table*}

\section{Transit Probabilities}{\label{app:probabilities}}

\begin{table*}
  \centering
  \caption{All non-zero transit probabilities of combinations of the Solar System planets, given as $\nicefrac{1}{N}$ and $\nicefrac{P}{P_{\earth}}$.}
  \rowcolors{2}{gray!25}{white}
    \begin{tabular}{lcc}
    \hline\hline
    Planets & Probability & Ratio to P$_{\earth}$ \\
    \hline
	Mercury & $1.2 {\times} 10^{-2}$ & $2.7$ \\
	Venus & $6.4 {\times} 10^{-3}$ & $1.4$ \\
	Earth & $4.6 {\times} 10^{-3}$ & $1.0$ \\
	Mars & $3.1 {\times} 10^{-3}$ & $6.6 {\times} 10^{-1}$ \\
	Jupiter & $8.1 {\times} 10^{-4}$ & $1.7 {\times} 10^{-1}$ \\
	Saturn & $4.5 {\times} 10^{-4}$ & $9.7 {\times} 10^{-2}$ \\
	Uranus & $2.3 {\times} 10^{-4}$ & $5.1 {\times} 10^{-2}$ \\
	Neptune & $1.5 {\times} 10^{-4}$ & $3.2 {\times} 10^{-2}$ \\
	Mercury, Venus & $6.7 {\times} 10^{-4}$ & $1.5 {\times} 10^{-1}$ \\
	Mercury, Earth & $3.0 {\times} 10^{-4}$ & $6.5 {\times} 10^{-2}$ \\
	Mercury, Mars & $2.7 {\times} 10^{-4}$ & $5.9 {\times} 10^{-2}$ \\
	Mercury, Jupiter & $5.9 {\times} 10^{-5}$ & $1.3 {\times} 10^{-2}$ \\
	Mercury, Saturn & $3.2 {\times} 10^{-5}$ & $7.0 {\times} 10^{-3}$ \\
	Mercury, Uranus & $1.7 {\times} 10^{-5}$ & $3.6 {\times} 10^{-3}$ \\
	Mercury, Neptune & $9.7 {\times} 10^{-6}$ & $2.1 {\times} 10^{-3}$ \\
	Venus, Earth & $3.2 {\times} 10^{-4}$ & $6.9 {\times} 10^{-2}$ \\
	Venus, Mars & $3.7 {\times} 10^{-4}$ & $8.0 {\times} 10^{-2}$ \\
	Venus, Jupiter & $8.3 {\times} 10^{-5}$ & $1.8 {\times} 10^{-2}$ \\
	Venus, Saturn & $5.1 {\times} 10^{-5}$ & $1.1 {\times} 10^{-2}$ \\
	Venus, Uranus & $2.1 {\times} 10^{-5}$ & $4.5 {\times} 10^{-3}$ \\
	Venus, Neptune & $1.2 {\times} 10^{-5}$ & $2.7 {\times} 10^{-3}$ \\
	Earth, Mars & $2.8 {\times} 10^{-4}$ & $6.1 {\times} 10^{-2}$ \\
	Earth, Jupiter & $1.0 {\times} 10^{-4}$ & $2.3 {\times} 10^{-2}$ \\
	Earth, Saturn & $3.0 {\times} 10^{-5}$ & $6.6 {\times} 10^{-3}$ \\
	Earth, Uranus & $5.2 {\times} 10^{-5}$ & $1.1 {\times} 10^{-2}$ \\
	Earth, Neptune & $1.4 {\times} 10^{-5}$ & $3.1 {\times} 10^{-3}$ \\
	Mars, Jupiter & $6.3 {\times} 10^{-5}$ & $1.4 {\times} 10^{-2}$ \\
	Mars, Saturn & $2.1 {\times} 10^{-5}$ & $4.6 {\times} 10^{-3}$ \\
	Mars, Uranus & $2.2 {\times} 10^{-5}$ & $4.8 {\times} 10^{-3}$ \\
	Mars, Neptune & $7.0 {\times} 10^{-6}$ & $1.5 {\times} 10^{-3}$ \\
	Jupiter, Saturn & $1.0 {\times} 10^{-5}$ & $2.3 {\times} 10^{-3}$ \\
	Jupiter, Uranus & $9.8 {\times} 10^{-6}$ & $2.1 {\times} 10^{-3}$ \\
	Jupiter, Neptune & $4.7 {\times} 10^{-6}$ & $1.0 {\times} 10^{-3}$ \\
	Saturn, Uranus & $2.0 {\times} 10^{-6}$ & $4.2 {\times} 10^{-4}$ \\
	Saturn, Neptune & $2.5 {\times} 10^{-6}$ & $5.4 {\times} 10^{-4}$ \\
	Uranus, Neptune & $8.4 {\times} 10^{-7}$ & $1.8 {\times} 10^{-4}$ \\
	Mercury, Venus, Saturn & $3.2 {\times} 10^{-5}$ & $7.0 {\times} 10^{-3}$ \\
	Mercury, Venus, Neptune & $1.6 {\times} 10^{-6}$ & $3.5 {\times} 10^{-4}$ \\
	Mercury, Earth, Mars & $2.1 {\times} 10^{-4}$ & $4.5 {\times} 10^{-2}$ \\
	Mercury, Earth, Uranus & $3.9 {\times} 10^{-8}$ & $8.4 {\times} 10^{-6}$ \\
	Mercury, Mars, Uranus & $4.6 {\times} 10^{-6}$ & $1.0 {\times} 10^{-3}$ \\
	Venus, Earth, Uranus & $2.1 {\times} 10^{-5}$ & $4.5 {\times} 10^{-3}$ \\
	Mars, Jupiter, Neptune & $2.2 {\times} 10^{-6}$ & $4.7 {\times} 10^{-4}$ \\
	Jupiter, Saturn, Uranus & $2.0 {\times} 10^{-6}$ & $4.2 {\times} 10^{-4}$ \\
    \hline
    \end{tabular}
  \label{tab:probabilities}
\end{table*}

\section{Exoplanets in Transit Zones}{\label{app:planetzones}}

\begin{table*}
	\centering
	\caption{68 confirmed and unconfirmed exoplanets that fall within one or more transit zones of the Solar System planets. Data from \url{http://exoplanet.eu}.}
	\rowcolors{2}{gray!25}{white}
	\begin{tabular}{llllr}
		\hline\hline
		Planet Identifier  & Status  & RA and Dec (Degrees)  & In Zones  & \multicolumn{1}{l}{Total} \\
		\hline
	    EPIC 211913977 b  & Confirmed  & 130.3441 18.9339  & Jupiter, Saturn, Uranus  & 3 \\
	    HATS-11 b  & Confirmed  & 289.4000 -22.3900  & Earth, Jupiter  & 2 \\
	    HD 181342 b  & Confirmed  & 290.2667 -23.6194  & Venus, Mars  & 2 \\
	    HD 50554 b  & Confirmed  & 103.6750 24.2456  & Venus, Mars  & 2 \\
	    K2-26 b  & Confirmed  & 94.2083 24.5958  & Venus, Mars  & 2 \\
	    11 Oph b  & Confirmed  & 245.6042 -24.0872  & Mercury  & 1 \\
	    1RXS 1609 b  & Confirmed  & 242.3750 -21.0828  & Earth  & 1 \\
	    2M 0441+23 b  & Confirmed  & 70.4375 23.0308  & Mars  & 1 \\
	    BD+20 594 b  & Confirmed  & 53.6500 20.5992  & Mercury  & 1 \\
	    EPIC 216468514 b  & Confirmed  & 284.9854 -22.2934  & Saturn  & 1 \\
	    GJ 785 b,c  & Confirmed  & 303.8208 -27.0331  & Mercury  & 1 \\
	    GJ 876 b,c,d,e  & Confirmed  & 343.3042 -14.2536  & Mercury  & 1 \\
	    GJ 876 f,g  & Unconfirmed  & 343.3042 -14.2536  & Mercury  & 1 \\
	    GSC 6214-210 b  & Confirmed  & 245.4792 -20.7186  & Venus  & 1 \\
	    HAT-P-50 b  & Confirmed  & 118.0625 28.1394  & Mercury  & 1 \\
	    HATS-3 b  & Confirmed  & 312.4583 -24.4289  & Mercury  & 1 \\
	    HD 164604 b  & Confirmed  & 270.7792 -28.5606  & Mercury  & 1 \\
	    HD 171238 b  & Confirmed  & 278.6833 -28.0722  & Mercury  & 1 \\
	    HD 179949 b  & Confirmed  & 288.8875 -24.1792  & Venus  & 1 \\
	    HD 204313 b,c,d  & Confirmed  & 322.0500 -21.7264  & Mercury  & 1 \\
	    HD 212771 b  & Confirmed  & 336.7625 -17.2636  & Mercury  & 1 \\
	    HD 222582 b  & Confirmed  & 355.4625 -5.9856  & Venus  & 1 \\
	    HD 283668 b  & Confirmed  & 66.9708 24.4447  & Mercury  & 1 \\
	    HD 284149 b  & Confirmed  & 61.6625 20.3031  & Venus  & 1 \\
	    HD 32963 b  & Confirmed  & 76.9833 26.3281  & Mercury  & 1 \\
	    HD 5319 b,c  & Confirmed  & 13.7542 0.7894  & Mercury  & 1 \\
	    HD 62509 b  & Confirmed  & 116.3250 28.0261  & Mercury  & 1 \\
	    HD 79498 b  & Confirmed  & 138.7875 23.3756  & Mercury  & 1 \\
	    HD 88133 b  & Confirmed  & 152.5292 18.1867  & Mercury  & 1 \\
	    HD 89307 b  & Confirmed  & 154.5875 12.6208  & Mars  & 1 \\
	    HIP 78530 b  & Confirmed  & 240.4792 -21.9803  & Mercury  & 1 \\
	    K2-14 b  & Confirmed  & 178.0570 2.5942  & Mars  & 1 \\
	    K2-15 b  & Confirmed  & 178.1108 4.2547  & Venus  & 1 \\
	    K2-17 b  & Confirmed  & 178.3298 6.4123  & Mercury  & 1 \\
	    K2-31 b  & Confirmed  & 245.4408 -23.5479  & Mercury  & 1 \\
	    LKCA 15 b  & Confirmed  & 69.8250 22.3508  & Earth  & 1 \\
	    MOA-2011-BLG-293L b  & Confirmed  & 268.9125 -28.4769  & Mercury  & 1 \\
	    MOA-2013-BLG-220L b  & Confirmed  & 270.9875 -28.4553  & Mercury  & 1 \\
	    OGLE-2011-BLG-0251L b  & Confirmed  & 264.5591 -27.1361  & Mercury  & 1 \\
	    OGLE-2012-BLG-0026L b,c & Confirmed  & 263.5792 -27.1428  & Mercury  & 1 \\
	    OGLE-2012-BLG-0358L b  & Confirmed  & 265.6958 -24.2611  & Venus  & 1 \\
	    OGLE-2012-BLG-0563L b  & Confirmed  & 271.4905 -27.7120  & Mercury  & 1 \\
	    OGLE-2013-BLG-0723L B b  & Unconfirmed  & 263.6708 -27.4481  & Mercury  & 1 \\
	    OGLE-2014-BLG-0124L b  & Confirmed  & 270.6208 -28.3964  & Mercury  & 1 \\
	    OGLE-2014-BLG-0257L b  & Confirmed  & 270.4500 -28.2619  & Mercury  & 1 \\
	    OGLE-2015-BLG-0051L b  & Confirmed  & 269.6625 -28.0317  & Mercury  & 1 \\
	    OGLE-2015-BLG-0954L b  & Confirmed  & 270.1833 -28.6608  & Mercury  & 1 \\
	    Pr 0201 b  & Confirmed  & 130.4333 20.2269  & Mars  & 1 \\
	    ROXs 12 b  & Confirmed  & 246.6042 -24.5433  & Mercury  & 1 \\
	    ROXs 42B b  & Confirmed  & 247.8125 -24.5456  & Mercury  & 1 \\
	    SR 12 AB c  & Confirmed  & 246.8333 -24.6944  & Mercury  & 1 \\
	    V830 Tau b  & Confirmed  & 68.2917 24.5619  & Mercury  & 1 \\
	    WASP-157 b  & Confirmed  & 201.6542 -8.3175  & Mars  & 1 \\
	    WASP-47 b,c,d,e  & Confirmed  & 331.2042 -12.0189  & Earth  & 1 \\
	    WASP-68 b  & Confirmed  & 305.0958 -19.3147  & Earth  & 1 \\
	    WD 1145+017 b  & Confirmed  & 177.1375 1.4831  & Earth  & 1 \\
		\hline 
	\end{tabular}
	\label{tab:planetzones}
\end{table*}

\label{lastpage}
\end{document}